\def\ref{\par\noindent\hang}

\def\spose#1{\hbox to 0pt{#1\hss}}
\def\approxlt{\mathrel{\spose{\lower 3pt\hbox{$\sim$}}
	\raise 2.0pt\hbox{$$<$$}}}
\def\approxgt{\mathrel{\spose{\lower 3pt\hbox{$\sim$}}
	\raise 2.0pt\hbox{$>$}}}

\def\multleft#1{\hbox to size{\vbox {\halign {\lft{##}\cr #1}}\hfill}\par}
\def\multright#1{\hbox to size{\vbox {\halign {\rt{##}\cr #1}}\hfill}\par}

\def\today{\ifcase\month\or January\or February\or March\or April\or May\or
      June\or July\or August\or September\or October\or November\or December\fi
      \space\number\day, \number\year}
\def\$<${\thinspace}
\def\s{\hbox{\phantom{5}}}	

\def\boxit#1{\vbox{\hrule\hbox{\vrule\kern3pt\vbox{\kern3pt
          #1 \kern3pt}\kern3pt\vrule}\hrule}}

\def\cm{{\rm\thinspace cm}}

\def\erg{{\rm\thinspace erg}}

\def\keV{{\rm\thinspace keV}}
\def\km{{\rm\thinspace km}}

\def\Lsun{\hbox{$\rm\thinspace L_{\odot}$}}

\def\Mpc{{\rm\thinspace Mpc}}
\def\Msun{\hbox{$\rm\thinspace M_{\odot}$}}

\def\s{{\rm\thinspace s}}

\def\ergpcmsqps{\hbox{$\erg\cm^{-2}\s^{-1}\,$}}

\def\ergps{\hbox{$\erg\s^{-1}\,$}}

\def\kmps{\hbox{$\km\s^{-1}\,$}}

\def\pcmsq{\hbox{$\cm^{-2}\,$}}

\def\ps{\hbox{$\s^{-1}\,$}}
\def\psqcm{\hbox{$\cm^{-2}\,$}}

\def\kmpspMpc{\hbox{$\kmps\Mpc^{-1}$}}

\documentstyle[psfig]{mn}
\begin{document}
\hsize=6truein

\title{Limits on the X-ray emission from several
hyperluminous IRAS galaxies}

\author[]
{\parbox[]{6.in} {R.J.~Wilman$^1$, A.C.~Fabian$^1$, R.M.~Cutri$^2$, 
C.S.~Crawford$^1$ and W.N.~Brandt$^3$ \\
\footnotesize
1. Institute of Astronomy, Madingley Road, Cambridge CB3 0HA \\
2. IPAC, Caltech, MS 100-22, Pasadena, CA 91125, USA \\
3. Department of Astronomy \& Astrophysics, Pennsylvania State University, 525 Davey Lab, University Park, PA 16802 USA \\ }}   

\maketitle

\begin{abstract}
We report long, pointed {\em ROSAT} HRI observations of the hyperluminous galaxies IRAS F00235+1024, F12514+1027, F14481+4454 and F14537+1950. Two of them are
optically classified as Seyfert-like. No X-ray sources are detected at the
positions of any of the objects, with a mean upper limit $L_{\rm{X}}/L_{\rm{Bol}} \simeq 2.3 \times 10^{-4}$. This indicates that any active nuclei are either atypically weak at X-ray wavelengths or obscured by column densities $N_{\rm{H}} > 10^{23} \pcmsq$. They differ markedly from `ordinary' Seyfert 2 galaxies, bearing a closer resemblance in the soft X-ray band to composite Seyfert 2 galaxies or to some types of starburst.   
\end{abstract}

\begin{keywords} 
galaxies:individual IRAS F00235+1024 -- galaxies:individual F12514+1027 -- galaxies:individual F14481+4454 -- galaxies:individual F14537+1950 -- infrared:galaxies -- X-rays:galaxies
\end{keywords}

\section{INTRODUCTION}
With bolometric luminosities in excess of $10^{13} \Lsun$ emitted mostly in
the mid- to far-infrared, hyperluminous {\em IRAS} galaxies are among the most
luminous objects in the Universe. The origin of the high luminosity,
however, remains uncertain, with massive starbursts and buried active nuclei
having both been implicated as radiation sources capable of powering thermal
reradiation by dust grains (see eg. Rowan-Robinson et al. 1993; Sanders \&
Mirabel 1996). Morphologically, many of these galaxies display the
signatures of recent interaction or merger events, either of which could
initiate bursts of star formation or drive gas into the central regions to
build/fuel active nuclei (AGN) (Sanders et al. 1988b). The model of Sanders et al. (1988a) describes how, following the collision of two gas-rich spirals, such systems evolve through being {\em IRAS} galaxies into optically-selected QSOs.

A number of hyperluminous galaxies possess Seyfert 2-like optical spectra,
with high-ionization emission-lines and strongly polarized continua
indicative of the presence of buried active nuclei [see eg. Hines et al.
(1995) for IRAS F15307+3252; Hines \& Wills (1993) and Kleinmann et al.
(1988) for IRAS P09104+4109]. If so, this would make possible a tentative
identification of hyperluminous galaxies as the long-sought class of type 2
quasars: Seyfert 2 galaxies with bolometric luminosities characteristic of
quasars.

X-ray observations exist for several hyperluminous galaxies and may provide
a useful means of discriminating between a starburst and an AGN, since the
latter are often prominent at these wavelengths. A source with strong iron K
line emission was detected by {\em ASCA} at the position of IRAS P09104+4109
(Fabian et al. 1994), favouring a model in which the observed X-rays are
scattered into our line of sight from a hidden quasar. A subsequent {\em ROSAT}
High Resolution Imager (HRI) image of the same source instead shows that the emission is actually dominated by a cooling flow (Fabian \& Crawford 1995), consistent with the observation that that object appears to reside at the core of a rich cluster of galaxies (Kleinmann et al. 1988). The hyperluminous galaxy IRAS F20460+1925 and the ultraluminous galaxy IRAS F23060+0505 are X-ray detected (Ogasaka et al. 1997 and Brandt et al. 1997, respectively) behind intrinsic column densities of $\sim 10^{22}\psqcm.$ The hyperluminous galaxy IRAS F10214+4724 was only marginally detected in X-rays [Lawrence et al. 1993: its luminosity is now known to be greatly enhanced by gravitational lensing (Eisenhardt et al. 1996)]. A 20-ks HRI observation of IRAS F15307+3252 (Fabian et al. 1996, hereafter F96) failed to detect any X-ray source at an upper limit of $\sim 4 \times 10^{43}\ergps$, representing less than $2 \times 10^{-4}$ of the bolometric luminosity and indicating either that the nucleus emits an anomalously small fraction of its total power in X-rays ($\ll L_{\rm{X}}/L_{\rm{Bol}} \sim 5$ per cent for a typical quasar) or that less than $0.4$ per cent of the nuclear X-ray flux is scattered into our line of sight.

\begin{table*}
\begin{center}

\caption{Summary of the properties of the four hyperluminous {\em IRAS} galaxies and the results of our analysis of the {\em ROSAT} HRI data}

\begin{tabular}{lllll} \hline
IRAS & F00235+1024 & F12514+1027 & F14481+4454 & F14537+1950 \\ \hline 
z & $0.575^{\ddagger}$ & $0.30^{\dagger}$ & $0.66^{\dagger}$ & $0.64^{\dagger}$
 \\ optical classification: & starburst & Seyfert 2 & Seyfert 2 & starburst \\
Galactic column density $N_{\rm{H}}:10^{20}\pcmsq$ & 5.1 & 1.73 & 1.72 & 2.81 \\
observation dates: & 1995 December 1 & 1996 July 11-14 & 1996 December 25 & 1996 August 9-12 \\
& 1996 July 2 & 1997 June 16 & 1997 July 14 \\ 
exposure time: s & 60365 & 20398 & 26883 & 17970 \\ 
count rate: $10^{-4}$ ct $\ps$ & $<0.70$ & $<2.40$ & $<1.72$& $<1.99$ \\$\Im$: $10^{-14} \ergpcmsqps$ & $<0.50$ & $<1.21$ & $<0.87$ & $<1.17$ \\ 
$L_{\rm{X}}: 10^{43} \ergps$ & $<0.87$ & $<0.54$ & $<2.07$ & $<2.61$ \\ 
$L_{\rm{Bol}}: 10^{13} \Lsun$ & $1.5^{\ddagger}$ & $1.0^{\dagger}$ & $2.0^{\dagger}$ & $2.0^{\dagger}$ \\ 
$L_{\rm{X}}/L_{\rm{Bol}}$ & $<1.5 \times 10^{-4}$ & $<1.4 \times 10^{-4}$ & $<2.7 \times 10^{-4}$ & $<3.4\times 10^{-4}$ \\ 
$\Delta N_{\rm{H}}: 10^{23} \pcmsq$ & 1.8 & 1.0 & 1.5 & 1.5 \\
$A_{\rm{V}}^{\star}$: mag & 120 & 67 & 100 & 100 \\ 
$L_{\rm{SX}}/L_{\rm{Bol}}$ & $<8.96 \times 10^{-5}$ & $<7.81 \times 10^{-5}$ & $<2.09 \times 10^{-4}$ & $<2.41 \times 10^{-4}$ \\ \hline

\end{tabular}
\end{center}

$\dagger$ from Cutri et al. (in preparation); $\ddagger$ from McMahon et al. (in preparation) \\ 
$\star$ using the $N_{\rm{H}}:E(B-V)$ conversion of Bohlin, Savage and Drake (1978), with a dust/gas ratio equal to that of the Galaxy. \\
\end{table*}

The key to an improved understanding of the hyperluminous galaxy phenomenon
lies in the study of larger samples. To this end we report {\em ROSAT} HRI
observations of a further four such objects, three of which were identified
by Cutri et al. (in preparation) using deep optical/radio imaging of {\em IRAS}
colour-selected `warm' extragalactic objects; the chosen objects are the
most luminous of those recently found by this method and have redshifts
substantially below that of IRAS F15307+3252, thus enhancing their
detectability. Two of these sources, IRAS F12514+1027 and F14481+4454 have
optical spectra similar to Seyfert 2 galaxies, indicating at least the presence
of an active nucleus.  The third object, IRAS F14537+1950, exhibits
starburst characteristics in its optical spectrum. The fourth object in our
sample, IRAS F00235+1024, was discovered by McMahon et al. (in preparation) during a systematic search for ultraluminous {\em IRAS} galaxies with the Cambridge APM machine; its optical spectrum was subsequently found to be starburst-like. When plotted on the {\em IRAS} colour diagram of Rowan-Robinson \& Crawford
(1989), IRAS F12514+1027 and IRAS F14481+4454 fall within the region occupied
by the Seyfert galaxies (galaxies with infrared spectra dominated by
starburst or disk components lie elsewhere in the diagram). The other two
objects in our sample cannot be placed on the diagram, since their {\em IRAS}
photometric data are of poorer quality.

\section{ROSAT HRI OBSERVATIONS}

IRAS F00235+1024, F12514+1027, F14481+4454 and F14537+1950 were observed with the {\em ROSAT} (Tr\"{u}mper 1983) HRI (David et al. 1995) between the dates given in Table 1. No sources are detected at the optical positions of the objects. Upper limits on the count rates (calculated at the 90 per cent confidence level using the Bayesian method of Kraft et al. 1991) are shown in Table 1. Also shown are upper limits on the unabsorbed flux, $\Im$, and luminosity, $L_{\rm{X}}$, for each object in the intrinsic $0.1-2.4 \keV$ band, assuming a power law continuum of photon index 2. For comparison with a sample of starburst galaxies in section 3.2, a thermal bremsstralung source spectrum with $kT=1 \keV$ was used to calculate $L_{\rm{SX}}$, the luminosity in the intrinsic
$0.5-2.0 \keV$ band. Allowance was made for photoelectric absorption by the
Galaxy (with column densities $N_{\rm{H}}$ shown in the table) and a cosmology of $H_{0}=50 \kmpspMpc$ and $q_{0}=0.5$ has been adopted throughout.

\section{DISCUSSION}

\subsection{On active nuclei}

From the table, we see that the upper limits on the ratio
$L_{\rm{X}}/L_{\rm{Bol}}$ are quite similar for all of the objects, with a
mean of $\simeq 2.3 \times 10^{-4}$. If we adopt in the first instance the
assumption that the sources are powered by active nuclei, we can use this
figure to make a number of inferences about their properties. A similar
discussion was given in F96 for the case of IRAS F15307+3252 which was
observed to have $L_{\rm{X}}/L_{\rm{Bol}} < 2 \times 10^{-4}$ (a $3 \sigma$
upper limit). We show, in Fig.~1, $L_{\rm{X}}/L_{\rm{Bol}}$ plotted against
$L_{\rm{Bol}}$ for these sources and some representative AGN and starbursts. Denoted by the arrows in this figure are the upper and lower ends of the range of values of $L_{\rm{X}}/L_{\rm{IR}}$ for the sample of Seyfert 2 galaxies in Green et al. (1992). The latter authors provide {\em Einstein} Imaging Proportional Counter (IPC) (0.5 - 4.5 keV) X-ray luminosities corrected for Galactic absorption which we convert to (0.1 - 2.4 keV) luminosities ($L_{\rm{X}}$). Luminosities in the 8 - 1000 $\mu$m band ($L_{\rm{IR}}$) were calculated from the {\em IRAS} photometric data in Bonatto et al. (1997) according to the prescription of Perault (1987). In our sample we include only those Seyfert 2 galaxies that were detected in all four {\em IRAS} bands and with the {\em Einstein} IPC. We note that for Seyfert 2 galaxies $L_{\rm{IR}}$ may differ substantially from $L_{\rm{Bol}}$, but we have insufficient information upon which to base a computation of the latter quantity. Nevertheless, since $L_{\rm{IR}} \simeq L_{\rm{Bol}}$ for the hyperluminous galaxies, the comparison is still a useful one. The large extent of the range for the Seyfert 2 galaxies is, we believe, a consequence of the heterogeneous nature of the sample, comprising composite objects with active nuclei and starbursts, as well as `ordinary' Seyfert 2 galaxies without the latter component. (We are not aware of any study that compares the values of $L_{\rm{X}}/L_{\rm{Bol}}$ for the known Seyfert 2 sub-populations).

\begin{figure}
\centerline{\psfig{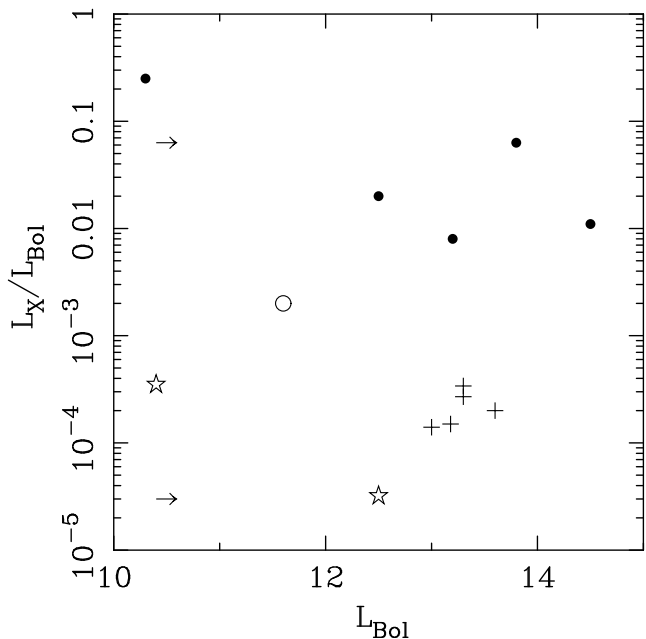}}
\caption{$L_{\rm{X}}/L_{\rm{Bol}}$ plotted against $L_{\rm{Bol}}.$ The
filled circles represent known AGN from MCG--6-30-15 on the left from
Reynolds et al. (1997), through IRAS F23060+0505 (Brandt et al. 1997) and F20460+1925 (Ogasaka et al. 1997) in the centre to PG 1634+706 and 1718+481 (Nandra et al. 1995) on the right. The crosses are the 90 per cent confidence level upper limits from the present data together with IRAS F15307+3252 (F96). The open circle is the Seyfert 2 galaxy NGC1068 (Wilson et al. 1992) and the stars
are (left to right) the starburst galaxies M82 (Kii et al. 1997, and Arp 220
(Kii et al. 1997). The arrows denote (for arbitrary $L_{\rm{Bol}}$) the upper and lower ends of the range for a sample of Sy 2 galaxies (see text for details). $L_{\rm X}$ for all objects has been corrected to an assumed 0.1--2.4~keV intrinsic power-law spectrum with a photon index of 2.}
\end{figure}

If being observed directly, it follows that the active nuclei are emitting
less than $0.02$ per cent of their power at soft X-ray wavelengths, which is
substantially below the figure of $5$ per cent for a typical quasar (F96).
Assuming that any active nuclei are intrinsically typical, the
existence of material with absorbent or scattering properties is thus
inferred. In the former case, ie. if nuclei are being observed in direct
light, quantities of material capable of depressing the soft X-ray fluxes by
minimum factors of between $\simeq 150$ (for IRAS F14537+1950) and $\simeq 360$ (for IRAS F12514+1027) are required. A proper treatment of red-shifted absorption using XSPEC (assuming an intrinsic power law continuum with a photon index of 2) demonstrated that this can be effected by intrinsic column densities
$\Delta N_{\rm{H}} >10^{23} \pcmsq$ (the inferred $\Delta N_{\rm{H}}$
for each object, along with the corresponding value of the visual extinction
$A_{\rm{V}}$, is shown in the table). This is very much less than $\Delta
N_{\rm{H}} \sim 1.5 \times 10^{24} \pcmsq$ at which the sources would become
optically thick to Compton scattering.

Alternatively, if the intrinsic column densities are large enough to block
all of the direct light, the results imply that on average less than 0.5 per cent of the soft X-ray flux is scattered into our line of sight (assuming a typical intrinsic value of 5 per cent for $L_{\rm X}/L_{\rm Bol}$; if the spread in the latter quantity is incorporated-- see Fig.~1-- the scattering fraction can be as high as $\sim3$ per cent). This is surprisingly small [cf. the $\simeq$ 10 per cent scattering fraction inferred in the optical for IRAS F15307+3252 by Hines et al. (1995)] and implies that any optical scattering medium is composed of dust (which does not scatter X-rays through large angles), not electrons. The hyperluminous galaxies are in this way different from nearby, lower luminosity Seyfert 2 galaxies such as NGC1068 where electron scattering plays an important role. Note too from Fig.~1 that they have a significantly lower value of $L_{\rm X}/L_{\rm Bol}$ than that of NGC1068, which is itself low for a (non-composite) Seyfert galaxy (see e.g. Awaki et al. 1997). This is perhaps an indication that even any scattered soft X-ray emission in our objects is absorbed.

It could instead be that intrinsically typical active nuclei are present but
that they account for less than 5 per cent of the total bolometric flux from
the objects (assuming a scattering fraction of 10 per cent). They may
therefore be high luminosity counterparts to composite starburst/Seyfert 2
galaxies, which, according to Ueno et al. (1997), fall toward the lower end of the Seyfert 2 range in Fig.~1, near the region occupied by our upper limits. Based upon the appearance of the optical spectrum, however, F96 provide reasons for believing this scenario to be unlikely for IRAS F15307+3252.

Indeed, given the large amounts of gas expected to be driven to the nucleus
of the galaxy during a merger and discoveries at low redshift that most
galaxies have central black holes (Magorrian et al. 1998), it would be
surprising if there were no luminous AGN in the objects observed. The
Eddington limit does however provide a restriction on the AGN luminosity
component, and requires a central mass exceeding $3-6\times 10^{8} \Msun$ in
order that the AGN component dominates a bolometric luminosity of
$L_{\rm{Edd}} = 1-2\times 10^{13}\Lsun$. If the merging galaxies had black
holes with masses ranging between that of our galaxy ($2.6\times 10^6\Msun$;
Eckart \& Genzel 1997) and M31 ($6\times 10^7\Msun$; Magorrian et al. 1998)
they could not then power the hyperluminous {\em IRAS} galaxies at the observed
rate. In other words, a merger between our galaxy and M31 could not produce
a hyperluminous AGN. (The Salpeter black hole growth timescale exceeds the
likely age of any merger.) A hyperluminous AGN requires a supermassive black
hole in the first place.

\subsection{On starbursts}

IRAS F00235+1024 and and IRAS F14537+1950 are classified as starbursts on
the basis of optical spectra (McMahon et al. in preparation and Cutri et al. in preparation, respectively). In order to assess the implications of the present soft X-ray data for a starburst interpretation for these objects, we compare our upper limits on the quantity $L_{\rm{SX}}/L_{\rm{Bol}}$ with the range of values given by Kii et al. (1997) for a sample of starburst galaxies. Our
definition of $L_{\rm{SX}}$ is motivated by the latter authors' use of the
soft X-ray {\em ASCA} band of $0.5-2.0 \keV$ and by the fact that the spectra of starburst galaxies are best described by thermal models. Kii et al. report
that $L_{\rm{SX}}/L_{\rm{Bol}}$ is of the order of $10^{-4}$ for starburst
galaxies, ranging from $3.2 \times 10^{-5}$ for Arp 220 [an ultraluminous
starburst galaxy the {\em ASCA} spectrum of which is analysed by Iwasawa (1998)] to $3.2 \times 10^{-4}$ for M82. It can be seen from Table 1 that the 90 per cent upper limits for the objects in our sample fall within this range and are thus not wholly inconsistent with a starburst origin for the soft X-ray emission. Our data suggest that, if starbursts do exist in these objects, they are more likely to resemble that in Arp 220 than that in M82.

\section{CONCLUSIONS}

We have reported the non-detection at soft X-ray wavelengths of four
hyperluminous galaxies. This result extends to a larger sample that found by F96 for IRAS F15307+3252, thus enabling constraints to be placed upon the properties of some of the most luminous objects in the Universe.

Two of the objects show evidence for an active nucleus in their optical
spectra. These AGN must be either anomalously weak at X-ray wavelengths or
obscured by column densities $N_{\rm H}>10^{23}\psqcm$. Any optical
scattering medium must be composed of dust, not electrons. The upper limits fall near the lower end of the range for a sample of Seyfert 2 galaxies, close to the region where objects of the latter type are thought to be of a composite nature (possessing starburst and active nucleus components). 

By comparison with {\em ASCA} observations of the soft X-ray fluxes from a sample of local starburst galaxies, we find that our {\em ROSAT} observations are not inconsistent with the presence of starbursts in our objects. For IRAS
F00235+1024 and F14537+1950, this is at least in line with expectations based
upon the appearance of their optical spectra.

\section*{ACKNOWLEDGEMENTS} We thank Richard McMahon for advance
information about IRAS F00235+1024. ACF and CSC thank the Royal Society for
support, and RJW acknowledges support from the PPARC. RMC acknowledges the
support of the Jet Propulsion Laboratory, Caltech which is operated under
contract with NASA.

\end{document}